\documentclass[lettersize,journal]{IEEEtran}
\usepackage{amsmath,amsfonts}
\usepackage{hyperref}
\usepackage{algorithmic}
\usepackage{array}
\usepackage[caption=false,font=normalsize,labelfont=sf,textfont=sf]{subfig}
\usepackage{textcomp}
\usepackage{stfloats}
\usepackage{url}
\usepackage{verbatim}
\usepackage{graphicx}
\def\BibTeX{{\rm B\kern-.05em{\sc i\kern-.025em b}\kern-.08em
    T\kern-.1667em\lower.7ex\hbox{E}\kern-.125emX}}
\usepackage{balance}
\begin{document}
\title{Intelligent Degradation Monitoring in Lithium-ion Batteries via Discharge Incremental Capacity Feature Estimation}

\author{Amir Madmolilvand, 
	Farzaneh Abdollahi
	\thanks{Amir Madmolilvand and Farzaneh Abdollahi are with the Department of Electrical Engineering, Amirkabir University of Technology, Tehran, Iran (e-mail: amir.mmv@aut.ac.ir; f\_abdollahi@aut.ac.ir).}
}

\maketitle

\begin{abstract}
Accurate and timely detection of degradation in lithium-ion batteries is crucial to ensure safety, reliability, and longevity in high-demand applications such as electric vehicles and energy storage systems. Traditional incremental capacity (IC) analysis methods require low-current cycling for discharge measurements, limiting their practical use in real-time battery management. This paper proposes a novel neural network-based framework that predicts discharge IC features directly from charging signals, eliminating the need for low-current discharge. Trained on a comprehensive dataset of 53 battery cells cycled under diverse fast-charging protocols, the model demonstrates robust generalization ability, effectively estimating degradation indicators on unseen battery data. Among several architectures evaluated, the LSTM model provides the best balance of prediction accuracy and computational efficiency. The proposed approach enables real-time integration into Battery Management Systems (BMS), enhancing degradation monitoring without disrupting normal battery operation. This paper presents a study to bridge IC analysis with practical, fast-charging scenarios, marking a significant step towards intelligent and scalable battery health monitoring.
\end{abstract}


\section{Introduction}
\IEEEPARstart{L}{ithium-ion} batteries (LIBs) are the predominant energy storage solution in contemporary applications; however, their long-term performance and safety are inherently linked to complex degradation mechanisms~\cite{kabir2017degradation, birkl2017degradation}. LIBs offer high energy density and extended cycle life (typically 1,000 to 6,000 cycles). On the other hand, they are prone to several limitations that remain a focus of ongoing researches ~\cite{betz2019theoretical, duffner2021post}.  Most of these limitations are further highlighted by concerns related to safety, cost, and environmental impact, prompting exploration into alternative battery chemistries and advanced battery management strategies.
\par
Despite continuous technological advancements, LIB systems encounter persistent degradation and performance-related issues ~\cite{shao2023novel,wang2019review,zheng2024quantitatively,hendricks2015failure}. This is particularly crucial in high-reliability applications such as electric vehicles (EVs), battery energy storage systems (BESS), and aerospace applications, which demand sustained performance under rigorous operational conditions. Accurate monitoring and assessment of the operational state of LIBs, coupled with robust safeguards against potentially hazardous conditions, is therefore paramount. The degradation pathways in LIBs are multifaceted, resulting in a spectrum of aging patterns and failure modes.
Consequently, effective Battery Management Systems (BMS) necessitate a comprehensive understanding of these degradation phenomena to optimize system performance and longevity.
\par

Recent research are focused on developing innovative approaches for monitoring and mitigating LIB degradation.  Operando techniques, such as impedance thickness methods, have been explored for real-time monitoring of lithium plating, utilizing impedance measurements and thickness monitoring, complemented by post-mortem analyses like Mass Spectrometry Titration (MST) and Scanning Electron Microscopy (SEM)~\cite{lin2024unveiling}. These approaches have identified critical degradation pathways. Similarly, studies employing Electrochemical Impedance Spectroscopy (EIS) on Nickel-Cobalt-Aluminum-Oxide (NCA) cells during cyclic overcharging have identified Loss of Conductivity (LoC), Loss of Lithium Inventory (LLI), and Loss of Active Material (LAM) as key degradation modes, with Incremental Capacity-Differential Voltage (IC-DV) analysis proposed as a diagnostic strategy~\cite{zhang2022degradation}. Further investigations into lithium plating mechanisms have integrated experimental methods with finite element modeling, using in-situ measurements, post-mortem analyses, and two-dimensional modeling, though practical implementation is constrained by the invasive nature of reference electrode methods and the additional cost of heat signal monitoring equipment ~\cite{mei2021understanding}.
\par
Model-based Differential Voltage Analysis (DVA) methods have been developed for Lithium Deposition (LD) detection, focusing on enhancing sensitivity and reliability. These methods have been extended to impedance relaxation techniques for adaptive charging control, minimizing capacity fade and reducing LAM and LLI~\cite{katzer2021model, katzer2023adaptive}. The integration of deep learning methodologies, such as the ICFormer model, has shown promise in leveraging incremental capacity (IC) curves to identify degradation modes and predict capacity degradation trajectories~\cite{costa2024icformer}. Furthermore, studies employing electrochemical and physical characterization techniques have quantitatively assessed the contributions of degradation mechanisms under extreme fast charging (XFC), correlating capacity fade and impedance growth with LLI and cathode material loss~\cite{yang2023quantitative}. Machine learning frameworks have been developed to differentiate lithium plating from solid electrolyte interphase (SEI) growth, leveraging electrochemical signatures such as capacity loss and Coulombic efficiency~\cite{chen2021machine}, and surface-level degradation analysis using Xenon ion (\({Xe}^+\)) plasma focused ion beam (PFIB) milling has provided insights into deactivation and parasitic reactions~\cite{yao2022degradation}. Acoustic ultrasound techniques have also been explored for lithium plating detection~\cite{bommier2020operando}.

%%~\cite{fly2020rate}, increasing the discharge rate tends to smooth the IC curve
\par
As previously mentioned, many studies have utilized IC analysis and differential voltage analysis in their approaches. However, these methods typically require the battery to be charged or discharged under low currents. The recommended current for accurate IC analysis is investigated in~\cite{fly2020rate}. In practical applications, cycling batteries at such low currents is infeasible due to prohibitive time requirements. Therefore, a novel approach that bridges the gap between IC analysis and real-world usage scenarios is needed.

In this paper, we propose employing a neural network to address this challenge. By enabling practical IC analysis as outlined, previous methods relying on IC for battery state-of-health (SOH) and capacity prediction~\cite{zeng2024lithium, lai2020combining}, or battery diagnosis and degradation estimation~\cite{costa2024icformer} will be more viable under operational conditions.
The IC curve can be obtained during both charge and discharge. However, increasing the current in either process leads to a smoothing effect in the IC curve, making the extraction of discriminative features more difficult. This limitation significantly impacts the applicability of charge IC analysis, particularly in real-world scenarios where fast charging is necessary. For discharge IC, if a dataset contains batteries charged with fast protocols but discharged at sufficiently low constant currents to produce informative IC curves, a neural network can be trained to map typical charging signals (e.g., voltage and current) to discharge IC features.
Despite extensive methodological efforts, challenges remain in practical deployment, time-intensive testing, and bridging the gap between laboratory and real-world conditions. Motivated by these challenges, the present study aims to quantify lithium-ion battery degradation during charging by employing a neural network that uses charging voltage and capacity signals as inputs to predict discharge incremental capacity features. This approach eliminates the need for actual battery discharge in batteries under normal operation (i.e., connected to a circuit and in use), thereby overcoming traditional constraints. By monitoring changes in the predicted discharge IC characteristics across successive cycles, degradation modes and faults in lithium-ion batteries can be effectively detected. The network is trained on historical data consisting of various fast-charging C-rates and profiles, and can be integrated into Battery Management Systems to provide real-time degradation detection without requiring low-current cycling. To the best of our knowledge, this study is the first to predict discharge incremental capacity features solely from charging data.

The key contributions of this work are summarized as follows:
\begin{itemize}
	\item A novel approach is presented to estimate the characteristics of the discharge incremental capacity curve based solely on charging data, enabling practical applicability in real-world scenarios without the need for low-current cycling.
	\item A neural network framework is proposed, extensively trained and validated on a large dataset of 53 NMC/graphite cells cycled under various fast charging protocols, demonstrating strong generalization capabilities across diverse conditions.
	\item It is shown that the proposed method can reliably extract diagnostic features indicative of battery degradation and faults, facilitating enhanced battery management and safety.
	\item It is validated that the model can be integrated into Battery Management Systems for real-time, non-intrusive degradation monitoring that circumvents traditional incremental capacity analysis limitations.
\end{itemize}

The remainder of this article is organized as follows: Section~\ref{sec:method} details the proposed methodology, including the dataset description, data partitioning strategy, analysis of aging effects on IC features, the design of the neural network architectures, and training procedures. Section~\ref{sec:results} presents the experimental results and discussion, highlighting model performance, generalization capability, and comparative analyses against baseline models. Finally, Section~\ref{sec:conclusion} concludes the paper by summarizing key contributions and outlining potential directions for future research.

\section{Method Description}
\label{sec:method}
\subsection{Dataset}
\label{sec:method_Datatset}
Incremental capacity (IC) can be calculated for both charge and discharge processes. As explained in the introduction, we focus on discharge IC. Therefore, the training and validation datasets for our task must include batteries charged at various C-rates but discharged under a unified, low C-rate. Ideally, this small discharge current should be around \(C/20\). One of the closest publicly available datasets that meets these criteria is the ISU-ILCC dataset~\cite{thelen2023isu}. The smallest discharge C-rate available in this dataset is \(0.5C\), which we thus consider as the discharge IC current for our study. As noted in~\cite{fly2020rate}, increasing the discharge rate tends to smooth the IC curve, thereby obscuring subtle changes within the battery. Consequently, an IC curve obtained at a discharge current of \(0.5C\) retains valuable diagnostic information, and tracking changes in its features remains indicative of underlying battery degradation modes.

%% DS Partioning
\subsection{Dataset Partitioning}
\label{sec:DS_partitioning}
ISU-ILCC dataset~\cite{thelen2023isu} comprises 225 nickel-manganese-cobalt/graphite (NMC/Gr) cells and was designed to systematically investigate battery aging under a wide range of operating conditions, including varying charge rates, discharge rates, and depth of discharge. These varied conditions yield a broad distribution of end-of-life (EOL) times, enabling robust training and validation of models aimed at battery degradation estimation.

In the dataset, 52 batteries meet the specified criteria, having been discharged at a constant rate of \(0.5C\) while charged at various practical rates including \(1C\), \(2C\), and others. We refer to this subset as the IC cells, while the remaining batteries form the nonIC cells. The nonIC cells are not applicable for our task, as their discharge IC is not available for each cycle at the required current. However, the dataset also provides weekly reference performance tests (RPTs), during which cells are charged and discharged at \(0.5C\). We utilize these RPT data of the nonIC cells for final validation to determine the best-performing model, a topic discussed in detail later.

Importantly, we do not use the RPT data alongside the IC dataset for training because in the RPT tests both charging and discharging occur at \(0.5C\). In real-world scenarios, if an RPT can be performed, the discharge IC would already be directly available, obviating the need for a neural network prediction. Moreover, performing weekly RPTs requires temporarily extracting the battery from its working conditions, which limits their practicality.

Despite this, the nonIC dataset remains valuable since differences in discharge current between the IC and nonIC cells might induce distinct aging dynamics. Given available datasets, this represents as close an approximation as possible to real-world applications. We anticipate that the model trained on the IC dataset will generalize well to the nonIC RPT data. Note that the RPT data for IC cells are excluded from our analysis.

For statistical evaluation, we employ \(k\)-fold cross-validation with \(k=6\) on the 52 IC cells. In each fold, 20\% of the batteries (approximately 9 to 10 cells) are reserved for validation, while the remaining 80\% are used for training.

\subsection{Aging Effects on IC Features}
\begin{figure}[!t] % 'htbp' specifies placement options  
	\centering % Centers the image  
	\includegraphics[width=9cm]{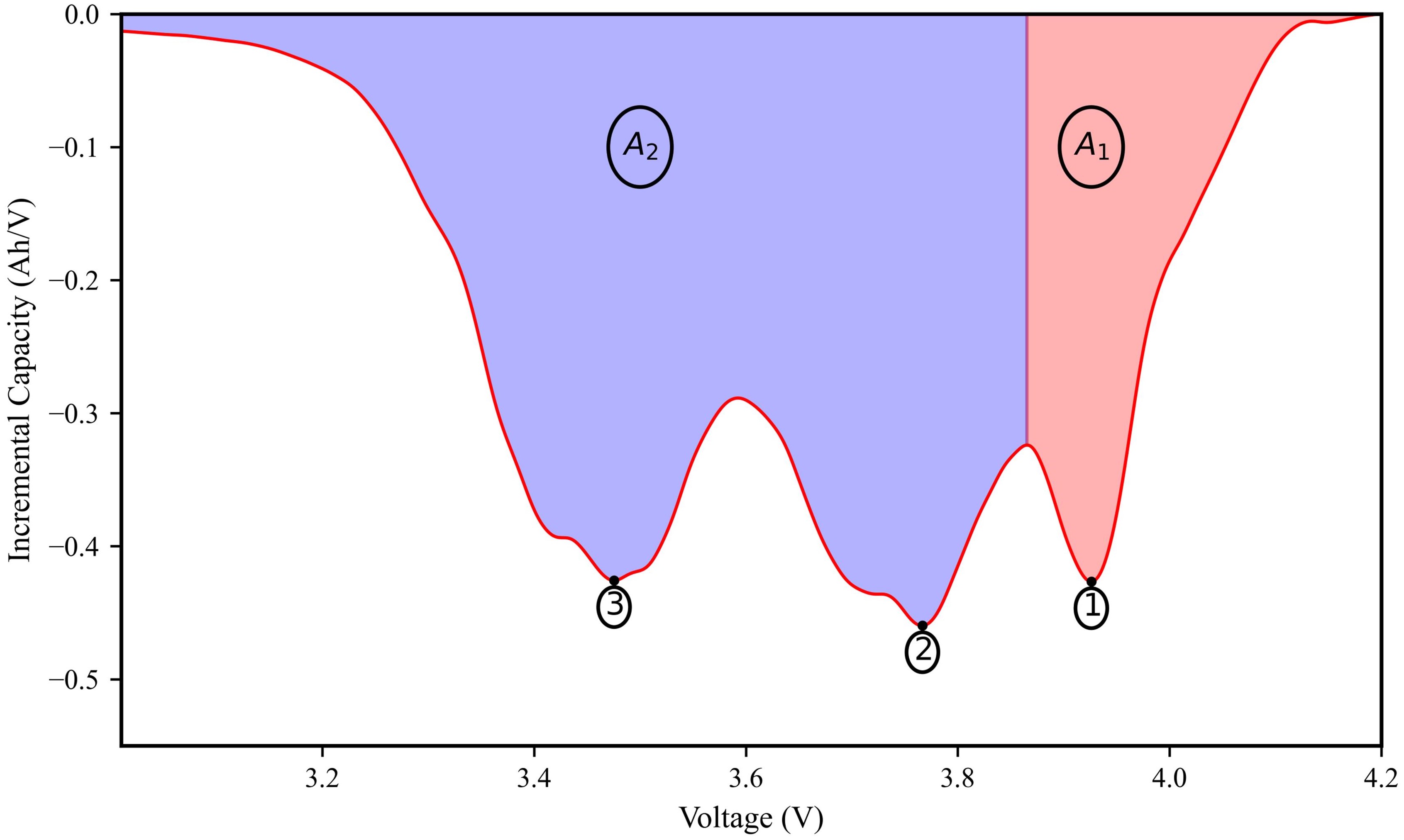} % Adjust width as needed  
	\caption{Schematic of Peak Numbers and Areas.
		The schematic representation illustrates the identification of peak numbers and areas.}
	\label{fig:Typical_IC} % Label for referencing the figure  
\end{figure}

\begin{figure*}[!t]
	\centering
	\includegraphics[width=0.8\textwidth]{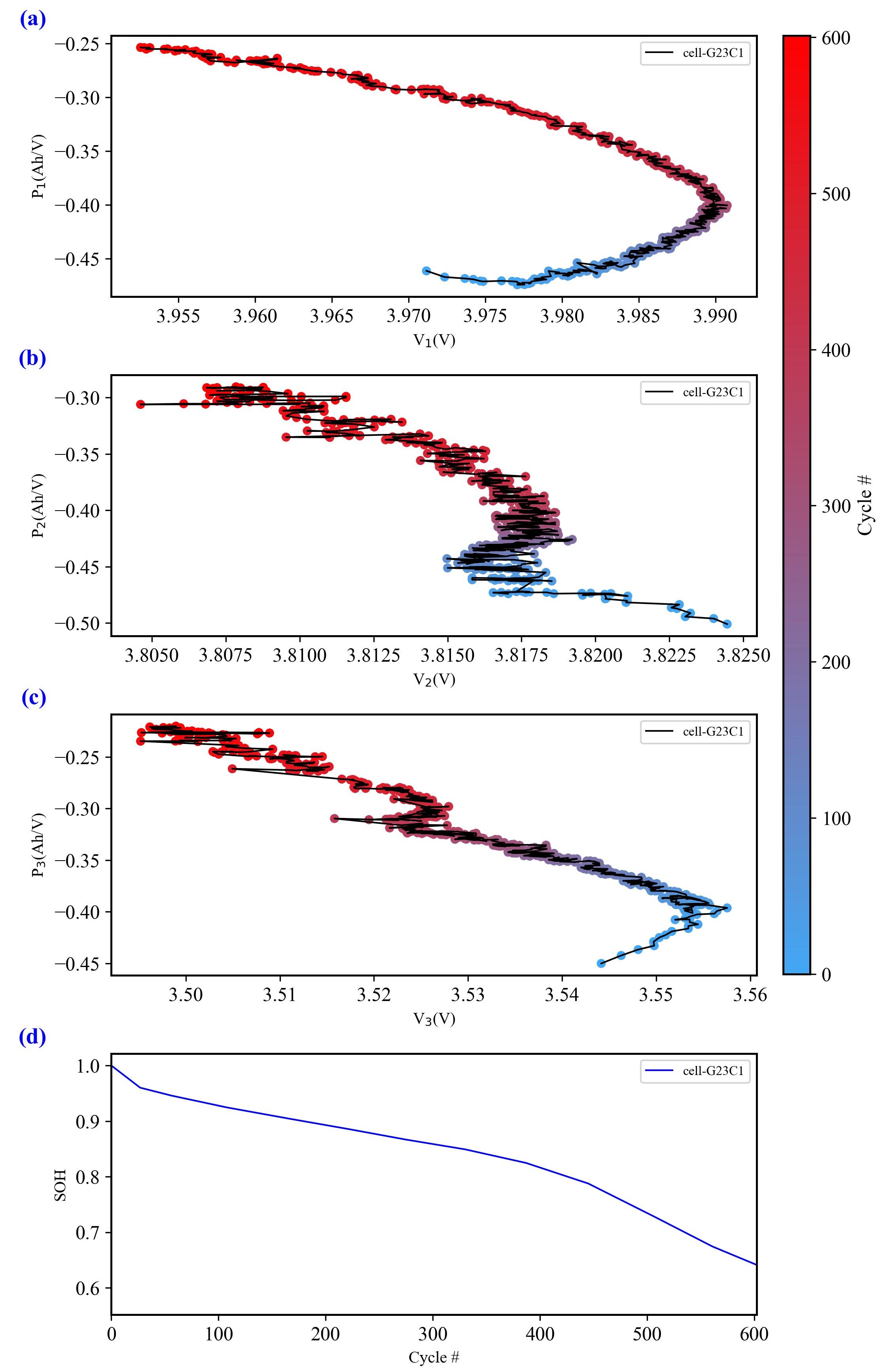} % Adjust width as needed  
	\caption{
		Evolution of incremental capacity (IC) peaks with respect to battery cycling and SOH. 
		The figure presents the evolution of IC peak locations, characterized by peak voltage and peak value, as a function of cycle number, indicated by the color bar. This is for battery cell G23C1.
		Subplots illustrate: 
		(a) the relationship between the value and voltage location of peak 1, 
		(b) the value and voltage location of peak 2, 
		(c) the value and voltage location of peak 3, and 
		(d) the SOH progression as a function of cycle number.
	}
	\label{fig:IC_features_deg}
\end{figure*}

To examine battery aging modes using IC, we refer to the IC features depicted in Figure~\ref{fig:Typical_IC}, adhering to the naming conventions established in~\cite{IC_2019}, where the significance of these features is thoroughly discussed. With one exception, the referenced study identifies five distinct peaks; however, three of these peaks are closely spaced in discharge IC curve. Due to the relatively higher discharge current in our dataset compared to the ideal \(C/20\), these three peaks effectively merge into a single peak, corresponding to peak 3 in our analysis. This represents a necessary compromise. Nevertheless, the variations of the three original peaks are similar across most degradation modes identifiable through discharge IC.
The voltage of peak \(i\) is \(V_i\) and the corresponding peak value is as follows:
\begin{equation}
	\label{eq:power_definition}
	P_i = \left.\frac{dQ}{dV}\right|_{V=V_i}
\end{equation}
\(A_1,A_2\) are the areas depicted in Figure~\ref{fig:Typical_IC}. Hence, the complete set of IC features is defined as
\begin{equation}
	S = \{P_1, P_2, P_3, V_1, V_2, V_3, A_1, A_2\}.
\end{equation}
%% Rewrite this 	
Figure~\ref{fig:IC_features_deg} illustrates the evolution of IC peak locations as a function of cycle number. Both the voltage and magnitude of each peak change noticeably as the battery ages. For instance, peak 1 exhibits an upward shift, which is indicative of loss of lithium inventory (LLI). Additionally, the figure demonstrates the degradation of battery state of health (SOH) over cycling. It is clear that degradation significantly influences each IC feature, and a strong correlation between SOH and IC characteristics is evident. These findings are consistent with previous work reported in~\cite{IC_2019}.

%% New 
To derive the required discharge IC curves, appropriate filtering and interpolation methods must be applied. Several practical challenges arise because directly extracting IC curves from raw measurement data results in significant noise. This noise level renders automated feature extraction nearly infeasible, as reliable identification of IC peaks and features typically requires manual expert intervention to accurately select relevant points. Therefore, robust preprocessing techniques are essential to obtain usable IC curves for subsequent analysis.

We used a Savitzky-Golay filter~\cite{wang2022state} with a window size corresponding to 10\% of the entire input signal and a polynomial order of 2 is applied to the data. This is followed by the application of a Gaussian filter~\cite{li2018quick} with \(\sigma = 10.0\). The discharge IC curve shown in Figure~\ref{fig:Typical_IC} is obtained using this procedure.
It is worth noting that certain IC peaks may disappear under extreme degradation conditions, which typically occur during the final cycles of battery life. Although these scenarios warrant further investigation, we discontinued the analysis at this stage, as extracting the IC features in such cases would require manual intervention.

% Plot Pi ,Vi and Ai with respect to cycle. 
\subsection{Problem Formulation}
Our goal is to use a neural network (NN) model that estimates the value of each discharge IC feature in \( S \) in each consecutive cycle using only charge signals, specifically cell voltage \( V_{\mathrm{charge}}(t) \) and capacity \( Q_{\mathrm{charge}}(t) \). It is very important to note that the input is from charge but the output is from discharge of the same cycle.

The input sequences for each data sample may vary in length, which can complicate the training process. To address this issue, we first interpolate each signal \(V_{\mathrm{charge}}(t)\) and \(Q_{\mathrm{charge}}(t)\) as functions of time \(t\), ensuring a uniform length across all data. This step is crucial for maintaining consistency and facilitating efficient processing by the model.
Thus, the network input data \(X\) can be represented as follows:
\begin{equation}
	\label{eq:inputX}
	X_{2\times n} = \begin{pmatrix}
		V_{\mathrm{charge}}^{\mathrm{T}}  \\
		Q_{\mathrm{charge}}^{\mathrm{T}}
	\end{pmatrix}_{2\times n} = \begin{pmatrix}
		V_1 & V_2 & \ldots & V_n  \\
		Q_1 & Q_2 & \ldots & Q_n
	\end{pmatrix}_{2\times n}
\end{equation}
where \(n\) is the selected length during time interpolation. For this work, we set \(n = 2000\).\\
For the model targets, we consider the discharge IC features as described in \(S\).
By combining these features, the network output will be as follows:
\begin{equation}
	\label{eq:OutPut_2}
	Y = \begin{pmatrix}
		V_1 \\
		V_2 \\
		V_3 \\
		P_1 \\
		P_2 \\
		P_3 \\
		A_1 \\
		A_2 \\
	\end{pmatrix}_{8 \times 1}
\end{equation}
Having mathematically formulated the problem, the next step is to develop a model that maps the input-output pairs with minimal loss. This involves training a network to learn the relationship between the charge signals and the corresponding discharge IC features, optimizing the model parameters to minimize the prediction error.

\subsection{Model Variations}
The goal of this study is to identify a model that achieves the best performance. However, the landscape of possible architectures is vast, each with numerous hyperparameters—such as activation functions, numbers of hidden layer nodes, and downsampling strategies that can significantly impact performance. 

We selected five widely used neural network architectures for evaluation: Multi-Layer Perceptron (MLP), Transformer~\cite{vaswani2017attention}, Long Short-Term Memory (LSTM)~\cite{hochreiter1997long}, bidirectional LSTM (BiLSTM)~\cite{schuster1997bidirectional}, and Convolutional Neural Network (CNN). Although other machine learning models, such as decision trees, may also perform well, this study focuses exclusively on neural networks.

Table~\ref{tab:ModelVariationParams} summarizes the number of trainable parameters for each model variation. To facilitate a fair comparison, we endeavored to keep the number of trainable parameters similar across all architectures. We conducted extensive hyperparameter tuning, and the architectures presented here represent the best-performing configurations discovered. It is worth noting that even better architectures may exist, but these models provide a robust baseline for comparison.

\begin{itemize}
	\item \textbf{LSTM}:
	The long short-term memory (LSTM) network employed in this study consists of two stacked LSTM layers, each with a hidden size of 32 units. The model is designed as a unidirectional recurrent architecture without dropout regularization. The LSTM outputs are followed by three fully connected layers with intermediate dimensions of 32, each activated by Leaky-ReLU nonlinearities to introduce non-linearity and mitigate potential vanishing gradient issues. The model weights are initialized using Xavier uniform initialization~\cite{glorot2010understanding} to facilitate stable convergence during training. During forward propagation, the hidden state from the last LSTM layer at the final time step is extracted and passed through the fully connected layers to generate the model output. This architecture balances complexity and expressiveness to capture temporal dependencies in the charge signals necessary to accurately estimate discharge IC features.
	\item \textbf{BiLSTM}:
	The bidirectional LSTM (biLSTM) network used in this study consists of two stacked LSTM layers, each with 32 hidden units. The LSTM layers are configured as bidirectional, enabling the model to capture temporal dependencies in both forward and backward directions. The output of the bidirectional LSTM, which has dimensionality twice the hidden size due to concatenation of forward and backward hidden states, is fed into a series of three fully connected layers with intermediate sizes of 32 units each. LeakyReLU activations are applied after the first two linear transformations to introduce non-linearity and improve model expressiveness. Weight initialization is performed using Xavier uniform initialization to promote stable and efficient training convergence. During the forward pass, the final hidden states from both directions of the last LSTM layer are concatenated and processed through the fully connected layers to produce the output feature vector. Most hyperparameters in our BiLSTM are similar to those of the unidirectional LSTM; however, the bidirectional nature leads to a larger number of trainable parameters.
	\item \textbf{MLP}:
	The multilayer perceptron (MLP) network employed in this study consists of three fully connected layers. The first two linear layers each map to 16 hidden units, followed by a final output layer of 8 units corresponding to the discharge IC features. LeakyReLU activation functions are applied after the first two linear transformations to introduce non-linearity and improve model generalization. Weight initialization uses Xavier uniform initialization to facilitate stable training convergence.
	\item \textbf{CNN}:
	The convolutional neural network (CNN) employed in this study consists of three 1D convolutional layers followed by three fully connected layers. The CNN layers have respective output channels of 32, 16, and 8, with kernel sizes of 9, 9, and 5, respectively. Each convolutional layer is followed by a LeakyReLU activation and max pooling operation with a kernel size of 5 to reduce spatial dimensionality while retaining important features. The fully connected layers have two intermediate layers of size 128 each and an output layer producing 8 discharge IC feature values. Weight initialization uses Xavier uniform initialization to facilitate stable training.
	\item \textbf{Transfomer}:
	The Transformer network employed in this study is designed to process 2-channel sensor data and consists of a 1D convolutional downsampling module, a positional encoding module, a multi-layer Transformer encoder, and a fully connected output projection. The downsampling module reduces the input sequence length from 2000 to 250 through a Conv1D layer with stride 4, followed by a ReLU activation and max pooling with kernel size 2. The network employs Rectified Linear Unit (ReLU) activation in the CNN component, which extracts salient features from the input data \(X\). To incorporate information about the order of the sequence, positional encoding based on sinusoidal functions is applied to the downsampled sequence. This encoding injects unique, deterministic signals to represent each position using sine and cosine functions of different frequencies, enabling the model to capture relative and absolute position information effectively~\cite{vaswani2017attention}. The core Transformer encoder consists of six layers, each equipped with two attention heads and a feedforward network of dimension 64, regularized with dropout of 0.1. After encoding, an average pooling operation aggregates the sequence information, and a final dense layer with eight neurons projects the representation into the output feature space. Xavier uniform initialization is employed across all model parameters to ensure stable and efficient training convergence.
\end{itemize}

%%%%%%%%%%%%%%%%%%%%%%%%%%%%%%%%
%%%%%%%%%%%%%%%%%%%%%%%%%%%%%%%%
%%%%%%%%%%%%%%%%%%%%%%%%%%%%%%%%

\subsection{NN Training}
To effectively train the neural network models, we employ the mean absolute error (MAE) loss function, as defined in Equation~\ref{eq:L_Y}. The neural network models are trained using the backpropagation algorithm, which iteratively adjusts network weights to minimize the prediction error between estimated and actual discharge IC features.
\begin{equation}
	\label{eq:L_Y}
	L_Y = \frac{1}{k} \sum_{i=1}^{k} \lVert Y_i - \hat{Y}_i \rVert
\end{equation}

where \(Y_i\) represent the true output values for the \(i\)th sample, and \(\hat{Y}_i\) denote the corresponding estimated value by the network. The total number of samples is denoted by \(k\).
Table~\ref{tab:NN_hyperparameters2} presents the hyperparameters employed during the training. All network variations were trained using a consistent set of hyperparameters to enable a fair comparison across architectures. Specifically, the Adam optimizer was employed with a learning rate of 0.0001, a batch size of 64, and a maximum of 200 training epochs.
As it was thoroughly explained in section~\ref{sec:DS_partitioning}, each model was trained with the same training and validation battery cells for each fold. 

\begin{table}[htbp]
\begin{center}
	\caption{Hyperparameters employed during the training of all network variations.}
		\label{tab:NN_hyperparameters2}
	\begin{tabular}{| c | c |}
		\hline
		\textbf{Training Hyperparameter} & \textbf{Value} \\% [6pt]
		\hline
		Learning Rate (\(\eta\)) & 0.0001 \\% [2pt]
		Batch Size & 64 \\% [2pt]
		Number of Epochs & 200 \\ %[2pt]
		Optimizer & Adam \\ % [2pt]
		Loss Function & \(L_Y\)\\ % [2pt]
		\hline
	\end{tabular}
\end{center}
\end{table}

\begin{figure*}[!t]
	\centering
	\includegraphics[width=\textwidth]{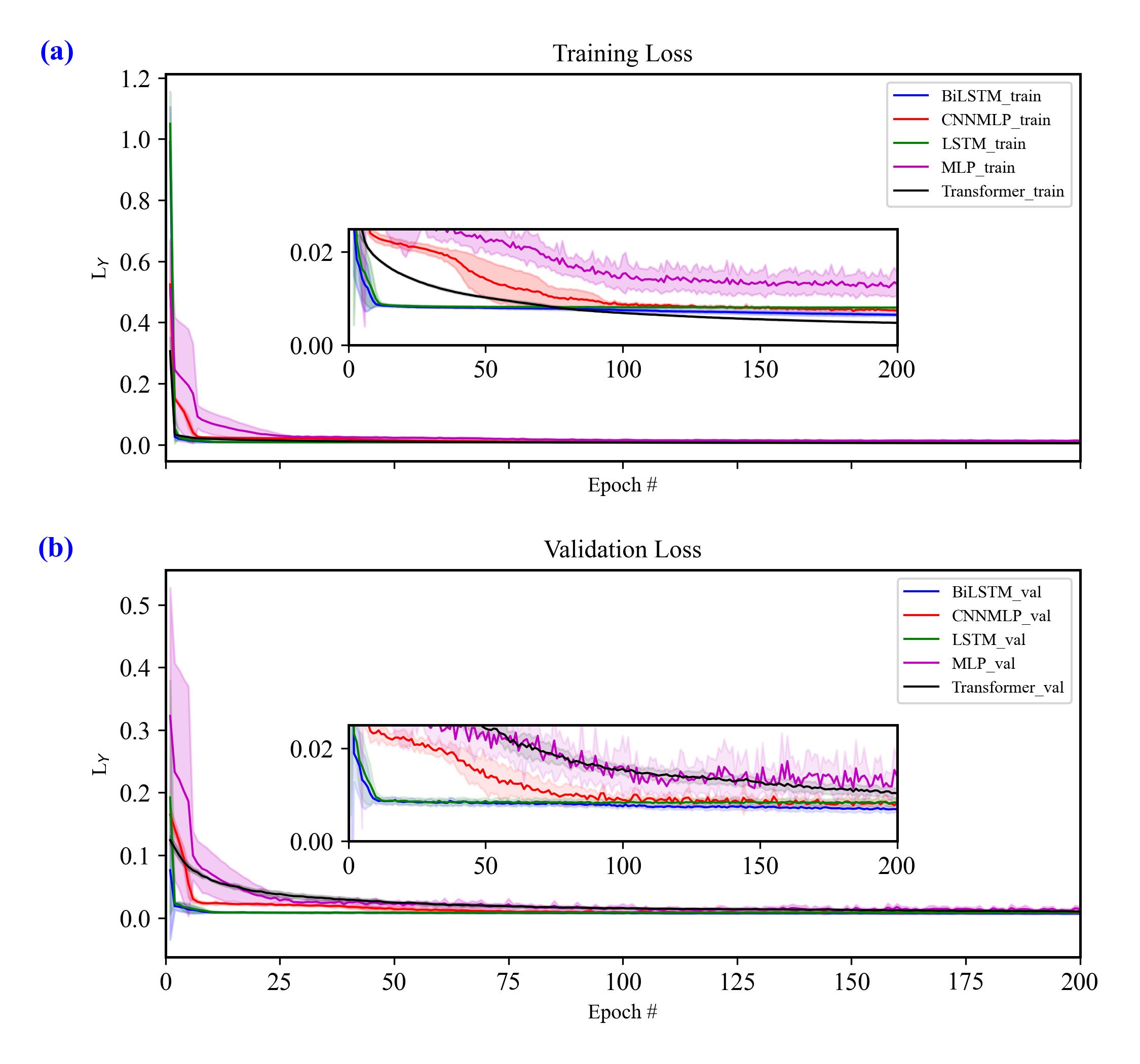}
	\caption{
		Training and validation loss \(L_Y\) progression over epochs for all model variations, averaged across all k-folds. Subplots display (a) training loss and (b) validation loss as functions of epoch for each dataset and model variant. The solid lines represent the mean loss across folds, while the shaded bands indicate the standard deviation, illustrating variability across folds for each model.
	}
	\label{fig:LossPlots}
\end{figure*}

\section{Results and Discussion}
\label{sec:results}

Figure~\ref{fig:LossPlots} illustrates the loss progression across epochs for all model variations. Almost all models converge to a small final loss value, with the notable exception of the MLP, which struggles to reach a stable minimum, as evidenced by the fluctuations in validation loss after epoch 150. The final loss values are summarized in Table~\ref{tab:KFoldLosses}.

It is important to note that the variation in each discharge IC feature with degradation is inherently small, as shown in Figure~\ref{fig:IC_features_deg}. Therefore, the observed small magnitudes of all final loss values are expected and reasonable. This underscores the need for a model that performs exceptionally well relative to others.

From the results in Table~\ref{tab:KFoldLosses}, the Transformer model achieves the lowest loss on training IC cells across all folds, with a mean final loss of \(4.75 \times 10^{-3} \pm 0.08 \times 10^{-3}\). However, the BiLSTM model attains the smallest validation loss, \(6.83 \times 10^{-3} \pm 0.60 \times 10^{-3}\), and the LSTM model performs best on non-IC cells, achieving a final loss of \(13.89 \times 10^{-3} \pm 0.24 \times 10^{-3}\).
Additionally, LSTM rivals BiLSTM on validation loss with \(8.39 \times 10^{-3} \pm 0.82 \times 10^{-3}\).
Table~\ref{tab:ModelVariationParams} presents the number of trainable parameters for each model variation. Among them, the LSTM model yields a favorable balance between performance and model complexity, offering competitive results with the smallest parameter count (15,432 trainable parameters). Based on these observations, it is reasonable to conclude that the LSTM model outperforms the other variations in this study with the best generalization ability.

\begin{table}[htbp]
\begin{center}
	\caption{Final mean MAE errors (\(L_Y\)) with standard deviations across all folds for different model variations. Results are reported separately for training and validation sets on IC cells, as well as final loss on nonIC cells.}
	\label{tab:KFoldLosses}

	\begin{tabular}{| c | c | c | c |}
		\hline
		\textbf{Model} & \textbf{Train \(\times 10^{-3}\)} & \textbf{Validation \(\times 10^{-3}\)} & \textbf{nonIC cells \(\times 10^{-3}\)} \\
\hline
		BiLSTM & $6.49 \pm 0.30$ & $6.83 \pm 0.60$ & $17.77 \pm 3.73$ \\
		LSTM & $8.04 \pm 0.13$ & $8.39 \pm 0.82$ & $13.89 \pm 0.24$ \\
		CNN & $7.38 \pm 0.23$ & $8.18 \pm 1.80$ & $16.81 \pm 1.05$ \\
		MLP & $13.14 \pm 2.62$ & $13.96 \pm 4.14$ & $21.53 \pm 3.28$ \\
		Transformer & $4.75 \pm 0.08$ & $10.30 \pm 1.16$ & $20.06 \pm 1.49$ \\
\hline
	\end{tabular}

\end{center}
\end{table}

\begin{table}[htbp]
\begin{center}
	\caption{Number of trainable parameters for each model variation evaluated in this study.}
	\label{tab:ModelVariationParams}
	\begin{tabular}{| c | c | c |}
\hline
		\textbf{Model} & \textbf{Number of Trainable Parameters} \\
\hline
		BiLSTM & 37,704 \\
		LSTM & 15,432 \\
		CNN & 37,888 \\
		MLP & 64,424 \\
		Transformer & 51,816 \\
\hline
	\end{tabular}
\end{center}
\end{table}

\section{Conclusion}
\label{sec:conclusion}

This paper proposed a novel neural network-based approach for intelligent degradation detection of Lithium-ion batteries through estimation of discharge incremental capacity features using only charging data. Our method overcomes the conventional limitations of incremental capacity analysis that require low-current cycling, thus enabling practical, real-time implementation within Battery Management Systems. 

Leveraging a comprehensive dataset of 53 NMC/graphite cells cycled under varied fast charging protocols, we employed a 6-fold cross-validation strategy to rigorously evaluate model performance across multiple training and validation splits. The trained models, particularly the LSTM architecture, demonstrated strong generalization capabilities, effectively capturing diagnostic features indicative of battery aging and degradation across diverse cycling conditions. This approach reliably extracted subtle degradation signatures from charging data alone, eliminating the need for disruptive and time-consuming discharge tests.

The results highlight the feasibility of integrating this data-driven framework into operational battery systems to enhance safety, extend battery life, and improve system efficiency. Future work will focus on expanding the methodology to diverse battery chemistries and formats, investigating robustness under varying operational and environmental scenarios, and implementing the trained model in embedded BMS hardware for on-line diagnostics and prognostics.

Overall, this study bridges the gap between laboratory incremental capacity diagnostics and real-world battery management applications, providing a practical and scalable solution for early fault detection and state-of-health estimation in Lithium-ion batteries.

\section*{Acknowledgments}
This research did not receive any specific grant from funding agencies in the public, commercial, or not-for-profit sectors. The data that support the findings of this study are openly available in the Iowa State University repository at \url{https://doi.org/10.25380/iastate.22582234.v2}~\cite{thelen2023isu}.

\bibliographystyle{IEEEtran}
\bibliography{references.bib}

@article{hochreiter1997long,
	title={Long short-term memory},
	author={Hochreiter, Sepp and Schmidhuber, J{\"u}rgen},
	journal={Neural computation},
	volume={9},
	number={8},
	pages={1735--1780},
	year={1997},
	publisher={MIT press}
}

@article{schuster1997bidirectional,
	title={Bidirectional recurrent neural networks},
	author={Schuster, Mike and Paliwal, Kuldip K},
	journal={IEEE transactions on Signal Processing},
	volume={45},
	number={11},
	pages={2673--2681},
	year={1997},
	publisher={Ieee}
}

@inproceedings{glorot2010understanding,
	title={Understanding the difficulty of training deep feedforward neural networks},
	author={Glorot, Xavier and Bengio, Yoshua},
	booktitle={Proceedings of the thirteenth international conference on artificial intelligence and statistics},
	pages={249--256},
	year={2010},
	organization={JMLR Workshop and Conference Proceedings}
}

@misc{thelen2023isu,
	author       = {Thelen, Adam and Li, Tingkai and Liu, Jinqiang and Tischer, Chad and Hu, Chao},
	title        = {ISU-ILCC Battery Aging Dataset [dataset]},
	year         = {2023},
	howpublished = {Dataset},
	publisher    = {Iowa State University},
	doi          = {10.25380/iastate.22582234.v2},
	url          = {https://doi.org/10.25380/iastate.22582234.v2}
}

@article{vaswani2017attention,
	title={Attention is all you need},
	author={Vaswani, Ashish and Shazeer, Noam and Parmar, Niki and Uszkoreit, Jakob and Jones, Llion and Gomez, Aidan N and Kaiser, {\L}ukasz and Polosukhin, Illia},
	journal={Advances in neural information processing systems},
	volume={30},
	year={2017}
}

@article{hendricks2015failure,
	title={A failure modes, mechanisms, and effects analysis (FMMEA) of lithium-ion batteries},
	author={Hendricks, Christopher and Williard, Nick and Mathew, Sony and Pecht, Michael},
	journal={Journal of Power Sources},
	volume={297},
	pages={113--120},
	year={2015},
	publisher={Elsevier}
}

@article{wang2019review,
	title={A review of lithium ion battery failure mechanisms and fire prevention strategies},
	author={Wang, Qingsong and Mao, Binbin and Stoliarov, Stanislav I and Sun, Jinhua},
	journal={Progress in Energy and Combustion Science},
	volume={73},
	pages={95--131},
	year={2019},
	publisher={Elsevier}
}

@article{shao2023novel,
	title={A novel method of discharge capacity prediction based on simplified electrochemical model-aging mechanism for lithium-ion batteries},
	author={Shao, Junya and Li, Junfu and Yuan, Weizhe and Dai, Changsong and Wang, Zhenbo and Zhao, Ming and Pecht, Michael},
	journal={Journal of Energy Storage},
	volume={61},
	pages={106788},
	year={2023},
	publisher={Elsevier}
}

@article{zeng2024lithium,
	title={Lithium-Ion Battery Capacity Estimation Based on Incremental Capacity Analysis and Deep Convolutional Neural Network},
	author={Zeng, Sibo and Chen, Sheng and Alkali, Babakalli},
	journal={Energies},
	volume={17},
	number={6},
	pages={1272},
	year={2024},
	publisher={MDPI}
}

@inproceedings{lai2020combining,
	title={Combining machine learning algorithms and an incremental capacity analysis on 18650 cell under different cycling temperature and SOC range},
	author={Lai, John and Chao, David and Wu, Alvin and Wang, Carl},
	booktitle={E3S Web of Conferences},
	volume={182},
	pages={03007},
	year={2020},
	organization={EDP Sciences}
}

@article{betz2019theoretical,
	title={Theoretical versus practical energy: a plea for more transparency in the energy calculation of different rechargeable battery systems},
	author={Betz, Johannes and Bieker, Georg and Meister, Paul and Placke, Tobias and Winter, Martin and Schmuch, Richard},
	journal={Advanced energy materials},
	volume={9},
	number={6},
	pages={1803170},
	year={2019},
	publisher={Wiley Online Library}
}

@article{duffner2021post,
	title={Post-lithium-ion battery cell production and its compatibility with lithium-ion cell production infrastructure},
	author={Duffner, Fabian and Kronemeyer, Niklas and T{\"u}bke, Jens and Leker, Jens and Winter, Martin and Schmuch, Richard},
	journal={Nature Energy},
	volume={6},
	number={2},
	pages={123--134},
	year={2021},
	publisher={Nature Publishing Group UK London}
}

@article{kabir2017degradation,
	title={Degradation mechanisms in Li-ion batteries: a state-of-the-art review},
	author={Kabir, MM and Demirocak, Dervis Emre},
	journal={International Journal of Energy Research},
	volume={41},
	number={14},
	pages={1963--1986},
	year={2017},
	publisher={Wiley Online Library}
}

@article{birkl2017degradation,
	title={Degradation diagnostics for lithium ion cells},
	author={Birkl, Christoph R and Roberts, Matthew R and McTurk, Euan and Bruce, Peter G and Howey, David A},
	journal={Journal of Power Sources},
	volume={341},
	pages={373--386},
	year={2017},
	publisher={Elsevier}
}

@article{IC_2019,
	title={Lithium-ion battery degradation indicators via incremental capacity analysis},
	author={Anse{\'a}n, David and Garc{\'\i}a, V{\'\i}ctor Manuel and Gonz{\'a}lez, Manuela and Blanco-Viejo, Cecilio and Viera, Juan Carlos and Pulido, Yoana Fern{\'a}ndez and S{\'a}nchez, Luciano},
	journal={IEEE Transactions on Industry Applications},
	volume={55},
	number={3},
	pages={2992--3002},
	year={2019},
	publisher={IEEE}
}

@article{fly2020rate,
	title={Rate dependency of incremental capacity analysis (dQ/dV) as a diagnostic tool for lithium-ion batteries},
	author={Fly, Ashley and Chen, Rui},
	journal={Journal of Energy Storage},
	volume={29},
	pages={101329},
	year={2020},
	publisher={Elsevier}
}

@article{lin2024unveiling,
	title={Unveiling the Three Stages of Li Plating and Dynamic Evolution Processes in Pouch C/LiFePO4 Batteries},
	author={Lin, Ying and Hu, Wenxuan and Ding, Meifang and Hu, Yonggang and Peng, Yufan and Liang, Jinding and Wei, Yimin and Fu, Ang and Lin, Jianrong and Yang, Yong},
	journal={Advanced Energy Materials},
	volume={14},
	number={36},
	pages={2400894},
	year={2024},
	publisher={Wiley Online Library}
}

@article{zhang2022degradation,
	title={Degradation characteristics investigation for lithium-ion cells with NCA cathode during overcharging},
	author={Zhang, Lei and Huang, Lvwei and Zhang, Zhaosheng and Wang, Zhenpo and Dorrell, David D},
	journal={Applied Energy},
	volume={327},
	pages={120026},
	year={2022},
	publisher={Elsevier}
}

@article{mei2021understanding,
	title={Understanding of Li-plating on graphite electrode: detection, quantification and mechanism revelation},
	author={Mei, Wenxin and Jiang, Lihua and Liang, Chen and Sun, Jinhua and Wang, Qingsong},
	journal={Energy Storage Materials},
	volume={41},
	pages={209--221},
	year={2021},
	publisher={Elsevier}
}

@article{katzer2021model,
	title={Model-based lithium deposition detection method using differential voltage analysis},
	author={Katzer, Felix and Jahn, Leonard and Hahn, Markus and Danzer, Michael A},
	journal={Journal of Power Sources},
	volume={512},
	pages={230449},
	year={2021},
	publisher={Elsevier}
}

@article{katzer2023adaptive,
	title={Adaptive fast charging control using impedance-based detection of lithium deposition},
	author={Katzer, Felix and M{\"o}{\ss}le, Patrick and Schamel, Maximilian and Danzer, Michael A},
	journal={Journal of Power Sources},
	volume={555},
	pages={232354},
	year={2023},
	publisher={Elsevier}
}

@article{costa2024icformer,
	title={ICFormer: A Deep Learning model for informed lithium-ion battery diagnosis and early knee detection},
	author={Costa, N and Anse{\'a}n, D and Dubarry, M and S{\'a}nchez, L},
	journal={Journal of Power Sources},
	volume={592},
	pages={233910},
	year={2024},
	publisher={Elsevier}
}

@article{yang2023quantitative,
	title={Quantitative analysis of origin of lithium inventory loss and interface evolution over extended fast charge aging in Li ion batteries},
	author={Yang, Zhenzhen and Tanim, Tanvir R and Liu, Haoyu and Bloom, Ira and Dufek, Eric J and Key, Baris and Ingram, Brian J},
	journal={ACS Applied Materials \& Interfaces},
	volume={15},
	number={31},
	pages={37410--37421},
	year={2023},
	publisher={ACS Publications}
}

@article{zheng2024quantitatively,
	title={Quantitatively detecting and characterizing metallic lithium in lithium-based batteries},
	author={Zheng, Zhi and Fang, Xue and Deng, Wei and Li, Peng and Zheng, Xiaobo and Zhang, Hang and Li, Lin and Chou, Shulei and Chen, Yuan and Tang, Yongbing and others},
	journal={Energy \& Environmental Science},
	year={2024},
	publisher={Royal Society of Chemistry}
}

@article{chen2021machine,
	title={A machine learning framework for early detection of lithium plating combining multiple physics-based electrochemical signatures},
	author={Chen, Bor-Rong and Kunz, M Ross and Tanim, Tanvir R and Dufek, Eric J},
	journal={Cell Reports Physical Science},
	volume={2},
	number={3},
	year={2021},
	publisher={Elsevier}
}

@article{yao2022degradation,
	title={Degradation Diagnostics from the Subsurface of lithium-ion battery electrodes},
	author={Yao, Xuhui and {\v{S}}amo{\v{r}}il, Tom{\'a}{\v{s}} and Dluho{\v{s}}, Ji{\v{r}}{\'\i} and Watts, John F and Du, Zhijia and Song, Bohang and Silva, S Ravi P and Sui, Tan and Zhao, Yunlong},
	journal={Energy \& Environmental Materials},
	volume={5},
	number={2},
	pages={662--669},
	year={2022},
	publisher={Wiley Online Library}
}

@article{bommier2020operando,
	title={In operando acoustic detection of lithium metal plating in commercial LiCoO2/graphite pouch cells},
	author={Bommier, Clement and Chang, Wesley and Lu, Yufang and Yeung, Justin and Davies, Greg and Mohr, Robert and Williams, Mateo and Steingart, Daniel},
	journal={Cell Reports Physical Science},
	volume={1},
	number={4},
	year={2020},
	publisher={Elsevier}
}

@article{li2018quick,
	title={A quick on-line state of health estimation method for Li-ion battery with incremental capacity curves processed by Gaussian filter},
	author={Li, Yi and Abdel-Monem, Mohamed and Gopalakrishnan, Rahul and Berecibar, Maitane and Nanini-Maury, Elise and Omar, Noshin and van den Bossche, Peter and Van Mierlo, Joeri},
	journal={Journal of Power Sources},
	volume={373},
	pages={40--53},
	year={2018},
	publisher={Elsevier}
}

@article{wang2022state,
	title={State of health estimation of lithium-ion battery in wide temperature range via temperature-aging coupling mechanism analysis},
	author={Wang, Limei and Qiao, Sibing and Lu, Dong and Zhang, Ying and Pan, Chaofeng and He, Zhigang and Zhao, Xiuliang and Wang, Ruochen},
	journal={Journal of Energy Storage},
	volume={47},
	pages={103618},
	year={2022},
	publisher={Elsevier}
}
\end{document}